\documentclass[12pt,nohyper,notoc]{JHEP}

\usepackage{amsmath,amssymb,cite,graphicx} 
	
\font\zfont = cmss10 
\newcommand\ZZ{\hbox{\zfont Z\kern-.4emZ}}
\def\inbar{\vrule height1.5ex width.4pt depth0pt}
\def\IC{\relax\hbox{\kern.25em$\inbar\kern-.3em{\rm C}$}}

\newcommand{\EQ}[1]{\begin{equation} #1 \end{equation}}

\newcommand{\SP}[1]{\begin{equation}\begin{split} #1 \end{split}\end{equation}}

\title{Graviton Propagators, Brane Bending and Bending of Light in
Theories with Quasi-Localized Gravity}

\author{Csaba Cs\'aki$^{a,}$\footnote{J. Robert Oppenheimer Fellow.}, 
Joshua Erlich$^a$ and Timothy J. Hollowood$^{a,b}$\\
$^a$Theory Division T-8, Los Alamos National Laboratory, Los Alamos,
NM 87545, USA\\
$^b$Department of Physics, University of Wales Swansea,
Swansea, SA2 8PP, UK\\

Email: {\tt csaki@lanl.gov, erlich@lanl.gov, pyth@skye.lanl.gov}}

\abstract{
We derive the graviton propagator on the brane for theories with
quasi-localized gravity. In these models the ordinary 4D graviton is
replaced by a resonance in the spectrum of massive Kaluza-Klein modes,
which can decay into the extra dimension. We find that the effects of
the extra polarization in the massive graviton propagator is exactly
cancelled by the bending of the brane due to the matter sources,
up to small corrections proportional to the width of the
resonance. Thus at intermediate scales 
the classic predictions of Einstein's gravity are reproduced in
these models to arbitrary precision.
}

\preprint{{\tt hep-th/0003020}}

\begin{document}

Following the work \cite{RS,RS2} 
of Randall and Sundrum (RS) there has been considerable
interest \cite{Nima,CS,junction,cvetic,Gremm,US,Gremm2,Linde,Verlinde,
skenderis,Kostas,Gubser,CHR,GT,GKR,Japanese,CGS2,other}
in the phenomenon of localization of gravity (for previous
relevant work see \cite{noncompact,oldCvetic}).
RS found
a solution to the five dimensional
Einstein equations in a background created by a single positive 
tension 3-brane and a negative bulk cosmological constant
which reproduces the effects of four-dimensional gravity
on the brane without the need to compactify the fifth
dimension. The reason for this is that the fluctuations of the 4D
metric are described by an ordinary quantum mechanical 
Schr\"odinger equation, where the shape of the QM potential 
resembles a volcano. This potential supports exactly one bound state
with zero energy, which can be identified with the 4D massless graviton,
since the wave functions of the massive continuum Kaluza-Klein (KK) 
modes are suppressed at the brane due to the tunneling through the 
potential barrier. Thus the effects of the KK modes are 
negligible for small energies, and at large distances ordinary
4D gravity is reproduced. 

Recently Gregory, Rubakov and Sibiryakov (GRS) found a brane model in which 
4D gravity is reproduced only at intermediate scales,
since at very small scales there are the same power-law corrections as
in the RS model, while
at very large scales the gravitational 
theory on the brane is again modified \cite{GRS}.\footnote{A similar
proposal has been made in \cite{Kogan}.} It has been explained,
that the reason behind this phenomenon (dubbed ``quasi-localization of
gravity'') is that the zero-mode of the QM system of these theories
becomes unstable and the exact bound state is replaced by a resonance of
zero mass in the continuum spectrum \cite{GRS,quasi,Dvali} (see also 
\cite{Witten}). 
As long as the lifetime of this 
resonance is large, there is a large region of intermediate scales
where an effective 4D Newton potential is reproduced. Once we get to 
large enough distances, the resonance will decay and therefore the
gravitational potential will be corrected.

It has been however suggested  in \cite{Dvali}, 
that even though the correct Newton potential is reproduced at
intermediate scales, one does not reproduce the results of
ordinary 4D Einstein gravity. The argument given in \cite{Dvali} for 
this is that in these models the resonant mode is a collective effect
of very light (but $m\neq 0$) KK modes. It is, however, well-known
that the $m\to 0$ limit of a massive graviton propagator does not reproduce the
massless graviton propagator, due to the fact that the number of polarizations
of the two fields do not match \cite{old}. Based on this discontinuity in
the graviton propagator as $m\to 0$, it was argued in \cite{Dvali}
that the predictions of theories with quasi-localization
of gravity would significantly differ from
those of ordinary gravity.  In particular the bending of light was predicted 
to be $\frac{3}{4}$ of the value in general relativity.

In this letter we show that this is in fact not the case. We find that up
to additional small corrections (which vanish in the limit in which the width 
of the resonance
$\Delta m\to 0$) the results of ordinary 4D gravity {\it are\/} reproduced on 
the brane at intermediate scales as in the RS scenario.  

The issue is that the
5D massless graviton propagator is given by
\EQ{
G_5(x,x')_{\mu\nu\rho\sigma}=\int\frac{d^5p}{(2\pi)^5}\,
\frac{e^{ip\cdot(x-x')}}{p^2}\big(\tfrac12g_{\mu\rho}g_{\nu\sigma}+
\tfrac12g_{\mu\sigma}g_{\nu\rho}-\tfrac13g_{\mu\nu}g_{\rho\sigma}\big)\ ,
}
while the 4D massless propagator is 
\EQ{
G_4(x,x')_{\mu\nu\rho\sigma}=\int\frac{d^4p}{(2\pi)^4}\,
\frac{e^{ip\cdot(x-x')}}{p^2}\big(\tfrac12g_{\mu\rho}g_{\nu\sigma}+
\tfrac12g_{\mu\sigma}g_{\nu\rho}-\tfrac12g_{\mu\nu}g_{\rho\sigma}\big)\ .
}
The difference in the tensor structure of 
these two propagators is caused by the presence of an
extra polarization state, a 4D scalar field, contained in 
the 5D graviton propagator.  Thus we see 
that the situation in theories
with localization and quasi-localization of gravity is not very different:
In the case of localized gravity the scalar
field must be included in the effective 4D theory, while in the case
of quasi-localized gravity the scalar is eaten by the massive 
graviton modes and appears as an additional graviton polarization.
One has to explain in both cases why the extra polarization state does not
eventually contribute to the propagator on the brane. 
Intuitively this is the case for the following reason: in brane models the 
stress tensor for a source on a brane has a vanishing 55 component;
thus we might na\"\i vely expect that one degree of freedom, namely
the scalar field, decouples 
from the sources on the brane and therefore the ordinary massless 
4D graviton propagator should be reproduced. It has been shown
in two beautiful papers by Garriga and Tanaka \cite{GT}, and by Giddings, Katz and
Randall \cite{GKR} that this is in fact the case. They have shown that once a 
source term on the brane is introduced the brane itself is bent in a frame in
which gravitational fluctuations are small.
The effect of the bending is such that it exactly compensates 
for the effect of the extra 4D scalar field contained in the 5D
graviton propagator,
and thus the ordinary 4D massless graviton propagator is reproduced in the
RS model. This hints that a similar situation may occur 
for theories with quasi-localized gravity, since the massive propagator
of the KK modes making up the resonance exactly the same tensor structure
as the massless graviton plus the scalar. We will show that the effect of
the brane bending will again cancel the effects of the extra polarization
in the massive propagator, up to corrections 
depending on the width of the resonance, which can be made
arbitrarily small by adjusting the parameters of the theory. Thus there
is no discontinuity in these models as $\Delta m\to 0$ and the results of
ordinary 4D Einstein theory are reproduced at intermediate scales
up to small corrections.

We now turn to the problem of constructing the propagators in a general
brane world described by the metric
\EQ{
ds^2=dy^2+e^{-A(y)}\eta_{\mu\nu}dx^\mu\,dx^\nu\ .
\label{nmetric}
}
We will take $A(y)$ to approach the
Randall-Sundrum form for $|y|\ll y_0$:
\EQ{
A(y)\to 2k|y| \ ,
}
while for $|y|\gg y_0$ the metric becomes flat:
\EQ{
A(y)\to {\rm constant}\ .
}
For instance, this is achieved in the GRS model \cite{GRS} by simply
patching anti-de Sitter space to flat space at some point $y=y_0$:
\EQ{
A(y)=\begin{cases} 2k|y| & |y|\leq y_0\\
2ky_0 & |y|\geq y_0\ .
\end{cases}
}
However, more generally one can imagine backgrounds that smoothly
interpolate between $AdS_5$ and flat space \cite{quasi}.
Since the metric approaches the RS metric for $|y|\ll y_0$, there is a
brane located at $y=0$ on which the matter fields will live.

We now study gravitational fluctuations in the background \eqref{nmetric}.
It is possible to choose a gauge for the gravitational fluctuations so
that they have the form 
\EQ{
ds^2=dy^2+e^{-A(y)}(\eta_{\mu\nu}+h_{\mu\nu})dx^\mu\,dx^\nu\ ,
\label{fluct}
}
where, in addition, we impose the transverse-traceless conditions
\EQ{
\partial^\nu h_{\mu\nu}=h^\mu_\mu \equiv h=0\ .
}
In this gauge, the transverse-traceless fluctuations satisfy the simple 
equation
\EQ{
\big(e^A\square^{(4)}+\partial_y^2-2A^\prime\,\partial_y\big)h_{\mu\nu}=0\ ,
\label{eom}
}
which is nothing but the scalar wave equation for each of the
components $h_{\mu\nu}$ in the background \eqref{nmetric}. At the
brane, we have the usual matching condition:
\EQ{
\partial_yh_{\mu\nu}\big|_{y=0}=0\ .
\label{mcz}
}
In the GRS model, there is an additional matching condition at the
point $y_0$, where $AdS_5$ is patched onto flat space:
\EQ{
\partial_yh_{\mu\nu}\big|_{y=y_0+}=\partial_yh_{\mu\nu}\big|_{y=y_0-}\ .
\label{jju}
}

Until now, the only matter sources in the theory were those that were
needed to produce the non-trivial background
\eqref{nmetric}.\footnote{The details of these background sources is not
important for our present discussion. We only need assume that the
sources vary linearly with a variation of the metric (see
\cite{US} for a discussion of this).} 
Now suppose we have additional sources corresponding to
matter on the brane.
The question is how this modifies the equations for the
fluctuations. It turns out the answer, as described in the 
papers by Garriga and Tanaka \cite{GT}, and by Giddings, Katz and Randall 
\cite{GKR}, is rather subtle but crucial for our analysis.  We closely follow
the former paper in what follows.

The point is that in the original coordinate system $(x^\mu,y)$, the presence of the
source on the brane actually bends the brane so that is no longer
situated at $y=0$. In order to investigate this effect, it is useful to
introduce a new coordinate system $(\bar x^\mu,\bar y)$ defined so
that the brane is at $\bar y=0$. We require that the metric in the new
coordinate system also has the form \eqref{fluct}, and this means that
the two coordinate systems must be related via
\EQ{
\bar y=y+\hat\xi^y(x)\ ,\qquad \bar
x^\mu=x^\mu+\hat\xi^\mu(x)-\partial^\mu
\hat\xi^y(x)\int^ydy'\,e^{A(y')}\ .
\label{trans}
}
Notice that the functions $\hat\xi^y(x)$ and $\hat\xi^\mu(x)$ do not
depend on $y$.
The relation between the metric fluctuations in the two coordinates
systems is then 
\EQ{
h_{\mu\nu}=\bar h_{\mu\nu}+\partial_\mu\hat\xi_\nu
+\partial_\nu\hat\xi_\mu-2\partial_\mu\partial_\nu
\hat\xi^y\int^ydy'\,e^{A(y')}
-\eta_{\mu\nu}\hat\xi^y\partial_yA\ .
\label{rbm}
}
At the moment, we have not specified the function $\hat\xi^y(x)$
and $\hat\xi^\mu(x)$; however, they will be determined
self-consistently, as we
shall see below.

The additional sources will modify the
matching condition at the brane \eqref{mcz}. In the new coordinate
system, where the brane is at $\bar y=0$, the Israel equations at the brane
give,
\EQ{
\partial_{\bar y}\bar h_{\mu\nu}\big|_{\bar y=0+}=
-\kappa^2\big(S_{\mu\nu}-\tfrac13\eta_{\mu\nu}S\big)
\ ,
\label{bbbe}
}
where $S_{\mu\nu}$ is given in terms of the matter stress tensor via
\EQ{
T_{\mu\nu}^{\rm brane}=S_{\mu\nu}(x)\delta(\bar{y})\ ,\qquad
T_{\bar{y}\bar{y}}^{\rm brane}=T_{\mu \bar{y}}^{\rm brane}=0.
}

Using the relation between the metric fluctuations \eqref{rbm} we can
now write down the boundary condition \eqref{bbbe} in the original
coordinates:\footnote{Notice that to leading order in the source we
can specify the following condition at $y=0$.} 
\EQ{
\partial_yh_{\mu\nu}\big|_{y=0+}=
-\kappa^2\big(S_{\mu\nu}-\tfrac13\eta_{\mu\nu}S\big)-
2\partial_\mu\partial_\nu\hat\xi^y
\ .
\label{nccg}
}
In the GRS model there is also the matching condition at $y_0$
\eqref{jju}. It is easy to show that this condition is simply
\EQ{
\partial_{\bar y}\bar h_{\mu\nu} \big|_{\bar y=\bar y_0+}
=\partial_{\bar y}\bar h_{\mu\nu}\big|_{\bar y=\bar y_0-}\ ,
\label{njju}
}
and so in the original coordinates \eqref{jju} remains unchanged.

The position of the brane in the original coordinate system is
$y=-\hat\xi^y(x)$ and so the 
condition \eqref{nccg} includes, via $\hat\xi^y(x)$, an effect from
the bending of
the brane. We now can now combine the equation of motion \eqref{eom} and the
boundary condition \eqref{nccg} to give
\EQ{
\big(e^A\square^{(4)}+\partial_y^2-2 A^\prime\,\partial_y\big)h_{\mu\nu}=-2\kappa^2
\Sigma_{\mu\nu}\delta(y)\
,
\label{neom}
}
where we have defined the effective source
\EQ{
\Sigma_{\mu\nu}=S_{\mu\nu}-\tfrac13\eta_{\mu\nu}S+\tfrac2{\kappa^2}
\partial_\mu\partial_\nu\hat\xi^y\ .
}
In order to determine $\hat\xi^y$, we now impose the fact that
$h_{\mu\nu}$ is traceless: $h=0$. The implies that the right-hand side
of \eqref{neom} itself must be traceless and so
\EQ{
\square^{(4)}\hat\xi^y=\tfrac{\kappa^2}6S\ ,
}
with solution
\EQ{
\hat\xi^y(x)=\frac{\kappa^2}6\int d^4x'\Delta_4(x,x')S(x')\ ,
}
where $\Delta_4(x,x')$ is the massless scalar Green's function for 4D Minkowski
space. If we now define the five-dimensional Green's function
\EQ{
\Big(e^A\square^{(4)}+\partial_y^2-2 A^\prime\,\partial_y\Big)\Delta_5(x,y;x',y')
=\delta^{(4)}(x-x')\delta(y-y')\ ,
}
then the fluctuation due to the source on the brane can be written
\EQ{
h_{\mu\nu}(x,y)=-2\kappa^2\int
d^4x'\,\Delta_5(x,y;x',0)\Sigma_{\mu\nu}(x')\ .
\label{jhg}
}

The final thing that we need to do is to transform the fluctuation
back into the coordinate system $(\bar x^\mu,\bar y)$ in which the
brane is straight. In order to have a simple final expression for the
fluctuation on the brane, we
can use the additional freedom 
in the transformation \eqref{trans}, present in $\hat\xi^\mu(x)$, to set,
\EQ{
\hat\xi_\mu(x)=\partial_\mu\Big(\tfrac1{2k}\hat\xi^y(x)-2\int
d^4x'\Delta_5(x,0;x',0)\hat\xi^y(x')\Big)\ .
}
Then from \eqref{rbm} and \eqref{jhg} we obtain our final result
for the fluctuation evaluated on the brane:
\EQ{
\bar h_{\mu\nu}(x,0)=-2\kappa^2\int d^4x'\Big\{\Delta_5(x,0;x',0)
\big(S_{\mu\nu}(x')-\tfrac13\eta_{\mu\nu}S(x')\big)-\tfrac{k}6\Delta_4(x,x')
\eta_{\mu\nu}S(x')\Big\} \ ,
\label{finres}
}
where $k=A^\prime(0)/2$.  Note that in a flat background the last term
in \eqref{finres} would be absent, and the theory is simply that of
five-dimensional gravity, as expected.\footnote{We thank John Terning for
raising this issue.}
The first term is exactly what one would have na\"\i vely expected,
while the second term arises from the effect of the bending of the brane 
\cite{GT,GKR}. 

Notice that the result \eqref{finres} is written in terms of the
scalar Green's function $\Delta_5(x,y;x',y')$ for the background
\eqref{nmetric} which allows us to make contact
with the auxiliary quantum mechanical system
of \cite{RS}. First of all, we relate the coordinate $z$ to 
 $y$ via
\EQ{
\frac{dz}{dy}=e^{A(y)/2}\ .
} 
The wavefunctions $\psi_m(z)$,
which satisfy the Schr\"odinger equation \cite{quasi}
\EQ{
-
\frac{d^2\psi_m}{dz^2}+\big(\tfrac9{16}A^{\prime2}-\tfrac34A^{\prime\prime}\big)
\psi_m=m^2\psi_m\ ,
\label{seq}
}
where $'\equiv d/dz$, and 
then give the Green's function that we need in \eqref{finres}. It is
expressed as a sum over the bound-states and an integral over the
continuum of KK modes:
\EQ{
\Delta_5(x,0;x',0)=\int \frac{d^4p}{(2\pi)^4}\,
e^{ip\cdot(x-x')}\Big\{\sum_m\frac{\psi_m(0)^2}{p^2+m^2}+
\int dm\,\frac{\psi_m(0)^2}{p^2+m^2}\Big\}\ .
}

The Schr\"odinger equation \eqref{seq} always admits a zero energy
state
\EQ{
\hat\psi_0(z)=N_0e^{-3A(z)/4}\ ,
}
which potentially describes the 4D graviton. In the RS model 
\EQ{
\hat\psi_0(z)=\frac{k^{1/2}}{(k|z|+1)^{3/2}}\ ,
}
is normalizable and so
\EQ{
\Delta_5(x,0;x',0)=k\Delta_4(x,x')+\cdots\ ,
}
where the ellipsis represents the  contribution from the KK modes. 
The contribution of the massless graviton fluctuation in
\eqref{finres} is then,
\EQ{
\bar h_{\mu\nu}(x,0)=-2\kappa^2k\int d^4x'\Delta_4(x,x')
\big(S_{\mu\nu}(x')-\tfrac13\eta_{\mu\nu}S(x')\big)+\cdots\ .
\label{hww}
}
However, this is {\it not\/} the usual propagator of a massless 4D
graviton. Fortunately, \eqref{hww} does not include the effect of the 
brane bending term in \eqref{finres}, which provides an additional
contribution that precisely has the effect of changing
$\tfrac13\to\tfrac12$ in \eqref{hww} and so yields the usual  
massless 4D graviton propagator.
So the brane bending effect is crucial even in the
original RS model \cite{GT,GKR}. 

In a quasi-localized gravity model, the transverse space is
asymptotically flat, {\it i.e.\/} $A(y)\to$ constant for $|y|\gg y_0$. 
In this case, the state $\hat\psi_0(z)$ is not
normalizable and there are only contributions to the Green's function from 
continuum modes. However, in \cite{quasi} we argued that if the scale $y_0$
is sufficiently large then $\hat\psi_0(z)$ appears as a sharp
resonance at the bottom of the continuum at $m=0$. In this case, 
we have approximately, for small $m$,
\EQ{
\psi_m(0)^2=\frac{{\cal A}}{m^2+\Delta m^2}+\cdots\ .
\label{bwf}
}
The height of the resonance is given by
\EQ{
\frac{{\cal A}}{\Delta m^2}=\hat\psi_0(0)^2=e^{-3(A(0)-A(\infty))/2}\ ,
\label{relo}
}
while the width, to leading order, can be determined by the fact that
as $y_0\to\infty$, the model should reduce to the RS model where
$\hat\psi_0(z)$ is normalizable. So as $\Delta m\to0$, \eqref{bwf}
should approximate $k\delta(m)$ giving
\EQ{
\frac{{\cal A}}{\Delta m}=\frac{2k}{\pi}\ .
\label{relt}
}
In the GRS model \eqref{relo} and \eqref{relt} give \cite{GRS} 
\EQ{
\Delta m=\frac{2k}{\pi}e^{-3ky_0}\ .
}

For small, but non-zero $\Delta m$, {\it i.e.\/} large $y_0\gg k^{-1}$, the resonance can
mimic the effect of the bound-state in the RS model. For
$|x-x'|\ll\Delta m^{-1}$, we can approximate the effect of the
resonance by a delta function and hence 
na\"\i vely we would expect a contribution to the graviton propagator
as in \eqref{hww}. However, just as in the RS model itself we have to
include the effect of the brane bending. This provides an additional
contribution which precisely has the effect of changing
$\tfrac13\to\tfrac12$ in \eqref{hww} and yields the usual massless 4D graviton
 propagator \cite{GT,GKR}.
So the effect of brane bending in the case of quasi-localized gravity
on the brane is exactly the same as for the RS model. We recover the
normal 4D graviton propagator in the regime where we can ignore the
contribution from the rest of the continuum, $|x-x'|\gg
k^{-1}$, and when we can approximate the resonance by a delta
function, $|x-x'|\ll\Delta m^{-1}$.
The effect of the finite width will give a calculable correction to
the graviton propagator that can be determined by
integrating over the shape of the resonance, giving
\SP{
\bar h_{\mu\nu}(x,0)=&-2\kappa^2k\int d^4x'\Big\{\Delta_4(x,x')
\big(S_{\mu\nu}(x')-\tfrac12\eta_{\mu\nu}S(x')\big)\\
&\qquad\qquad\qquad\qquad\qquad+
\Delta m\,\tilde\Delta_4(x,x')\big(S_{\mu\nu}(x')-\tfrac13\eta_{\mu\nu}S(x')\big)\Big\}
\ ,
}
where 
\EQ{
\tilde\Delta_4(x,x')=-\int
\frac{d^4p}{(2\pi)^4}\frac{e^{ip\cdot(x-x')}}{p^2(p+\Delta m)}\ .
}

We have studied theories in which gravity is quasi-localized to a brane.
In these
theories there is a massless graviton resonance which eventually decays
into the bulk, with the result of altering the very long distance behavior
of gravity.  It has been suggested that there are phenomenological
difficulties with such models as a result of the non-decoupling of massive
graviton polarizations in the massless limit.  In this letter we have
shown
that in fact such difficulties are absent.  
The reason is that the bending of the brane exactly compensates for the
effects of the extra polarization in the massive graviton 
propagator (just as it compensates for the effect of the massless 
scalar in the case of the RS theory). Thus the graviton propagator at 
intermediate distances will be equal to the massless propagator of the
Einstein theory (up to corrections that can be made arbitrarily small
by making the width of the resonance small).  Therefore, all classic 
predictions of general relativity including the bending of light around
the Sun and the precession rate of the orbit of Mercury are correctly
reproduced in these theories. 

Theories with quasi-localized gravity open exciting possibilities for
phenomenology.  Because gravity is modified at both very
short and very long distances in such models, there are phenomenological
consequences for both particle physics at high energies and cosmology at large
distances.  These consequences merit further investigation.

We would like to thank Tanmoy Bhattacharya 
and John Terning for several useful
conversations, and Gia Dvali, Ruth Gregory and Valery Rubakov for
comments on the manuscript and for sharing their views on these 
theories with us.
C.C. is an Oppenheimer Fellow at the  Los Alamos National Laboratory.
C.C., J.E. and T.J.H. are supported 
by the US Department of Energy under contract W-7405-ENG-36.

\end{document}